
\magnification=\magstep1
\hsize 32 pc
\vsize 42 pc
\baselineskip = 24 true pt
\def\cl{\centerline}
\def\vs {\vskip .5 true cm}
\cl {\bf A Simple Model for Maxwell's Demon Type}
\cl {\bf Information Engine}
\vs
\cl {\bf A.M. Jayannavar$^*$}
\cl { Institute of Physics, Sachivalaya Marg,,}
\cl {Bhubaneswar - 751005, India}
\vs
\noindent {\bf Abstract}

We have investigated the recently proposed self-consistent theory of
fluctuation - induced transport. In this framework the subsystem
under study is coupled to two independent baths at different
temperatures. In this non-equilibrium system one can extract energy
at the expense of increased entropy. This is a simple model of
Maxwell's demon engine that extracts work out of a non-equilibrium
bath by rectifying internal fluctuations. We point out the errors in
the earlier results. We have obtained an analytical expression
for the fluctuation - induced transport current in a non-equilibrium
state and various cases of physical interest have been elucidated.
\vskip 1.5 true cm
\noindent PACS Numbers : 05.60, +w, 05.40.+j, 87.10.+e.
\vskip 1.5 true cm
\noindent * e-mail address : jayan@iopb.ernet.in
\vfill
\eject
{}From thermodynamics, it is well-known that useful work cannot be
extracted from equilibrium fluctuations [1]. In a thermal equilibrium
state the principle of detailed balance ensures that no net particle
current can flow in the presence of external potential of arbitrary
shape. In contrast, in a non-equilibrium situation, where detailed
balance is lost, net current flow is possible, i.e., one can extract
energy at the expense of increased entropy. Several models have been
proposed recently in this direction [2-6]. The motion of a heavily
damped Brownian particle, in the presence of asymmetric static
external potential and under non-white or correlated fluctuations, is
a simple example of a non-equilibrium system. In such a system
induced current or directed motion appears, eventhough the average of
the driving fluctuations vanishes. It turns out that the preferred
direction of motion and the magnitude of induced current depends
sensitively on the shape of the potential as well as on the
statistics of the fluctuations. Moreover, to obtain induced current
the strength of non-white noise must exceed a certain minimum value
and the magnitude of induced current shows a  maximum value at an
intermediate value of the noise strength. These models [2-7] of
engines to obtain coherent response (or rectification) from unbiased
forcing come under the common denomination of ``thermal ratchets'' or
``fluctuation - induced transport systems''. The idea of thermal
ratchets  has been utilized recently for molecular separation
[8]. One of the major motivations of these studies comes from
molecular biophysics, where ratchet like mechanism is proposed to
explain unidirectional movement of macromolecules or molecular motors. This is
a
physical example of preferred directional motion of Brownian particles
(macromolecules) along periodic structures in the absence of obvious
driving potentials, such as chemical potential gradients or thermal gradients.

In a physically well motivated recent work [9], Millonas points out
that all the proposed earlier models [2-6] are basically
phenomenological in nature and no attempt has been made to formulate
the problem from first principles. Millonas in his treatment [9]
constructs a Maxwell's demon-type information engine that extracts
work from non-equilibrium bath and allows a rigorous
determination of
kinetics consistent with the underlying laws of physics. He
explicitly writes down a microscopic Hamiltonian including the
subsystem and two thermal baths at different temperatures. An
existing inequality of temperature can be exploited to do useful
work. After
eliminating bath variables one obtains non-linear Langevin equation
for the subsystem variable Q : namely
$$M\ddot Q + \Gamma(Q)\dot Q +  U' (Q) =\xi_A
(t)+\sqrt{f(Q)}\xi_B(t),\eqno{(1)}$$
where $\Gamma (Q) = (\Gamma_A+\Gamma_B f(Q)),\xi_A(t) $ and $\xi_B(t)$ are
two independent Gaussian white-noise fluctuating forces with
statistics given by
$$ <\xi_A (t) > = 0, \ < \xi_A(t)\xi_A(t')> = 2 \ \Gamma_A kT\delta
(t-t'),\eqno{(2.a)}$$
and
$$ <\xi_B (t) > = 0, \ < \xi_B(t)\xi_B(t')> = 2 \ \Gamma_B k\overline T\delta
(t-t'),\eqno{(2.b)}$$
T and $\overline T$ are temperatures of the two baths A and B,
respectively. Henceforth we set Boltzmann constant k to be unity. The
bath B is characterised by a space dependent friction coefficient
$\Gamma_B f(Q)$. In eq.(1) $U'(Q)$ is an external force
(correspondingly U(Q) is an external potential) and we have used
expression f(Q) for $[V'(Q)]^2$ of ref.[9]. The final calculations
in ref.[9] are done in an overdamped limit by simply neglecting
$\ddot Q$ term in eqn.(1). We have shown elsewhere that such a
procedure is incorrect [10]. The resulting Fokker-Planck equation
(eqn.(8) of ref.[9]), turns out to be inconsistent. For example, in an
equilibrium situation $(T =\overline T)$, when the potential U(Q) is
unbounded and positive, i.e., $U(Q)\rightarrow \infty$ as
$Q\rightarrow \pm\infty,$ the equilibrium distribution comes out to be
$$P_c(Q) = C\Gamma (Q) e^{-U(Q)/T},\eqno{(3)}$$
where C is the normalization constant. This distribution function is incorrect
because in equilibrium $P_e(Q)$ must have the form
$$ P_e(Q) = N  e^{-U(Q)/T},\eqno{(4)}$$
where N is the position independent normalization constant. Also, if we
set $\Gamma_A=0$ (i.e, the particle is coupled to a single bath B at
temperature $\overline T$) again the  equilibrium distribution of the
form of eqn.(3) is obtained. The reason for this inconsistency can be
traced back to the improper overdamped limit of the original Langevin
equation (eqn.(1)). For example in the absence of thermal bath A,
Sancho et al. [11] have shown that the correct ovedamped Langevin
equation should be
$$\Gamma_B f(Q)\dot Q = - U'(Q) - \overline T {[\sqrt{f(Q)}]'\over
\Gamma_B\sqrt{f(Q)}} +\sqrt{f(Q)}\xi_B(t).\eqno{(5)}$$
The above equation leads to the correct equilibrium distribution as
mentioned above. The error in the Millonas treatment [9] follows from
the incorrect overdamped limit, where he ignores a term like the
second term on the right hand side of eqn.(5).

In this brief report, we construct the correct Fokker-Planck equation
in the overdamped limit. We use this equation to study fluctuation -
induced transport in a system where the function U(Q) and f(Q) are
periodic under translation $Q\rightarrow Q +\lambda ,
(U(Q+\lambda)=U(Q)$ and $f(Q+\lambda) = f(Q))$. We obtain the correct
expression for the mean velocity $<\dot Q >$  and study several
special cases of physical interest.

Following ref.[11], after a straightforward algebra one can
readily obtain Fokker-Planck equation for the variable  Q(i.e, the evolution
equation
for the probability density P(Q,t)), in the overdamped limit and is
given by
\vfill
\eject

$${\partial P\over\partial t}= {\partial\over\partial Q}
{U'(Q)\over\Gamma(Q)} P + T \Gamma_A {\partial\over\partial Q}{1\over
\Gamma (Q)}{\partial\over\partial Q}{1\over\Gamma (Q)} P .$$
$$ +\overline T \Gamma_B {\partial\over \partial Q}{\sqrt{f(Q)}\over
\Gamma (Q)} {\partial\over \partial Q}{\sqrt{f(Q)}\over \Gamma (Q)} P
+ \overline T \Gamma_B {\partial\over \partial
Q}{[\sqrt{f(Q)}]'\sqrt{f(Q)} \over [\Gamma (Q)]^2}P . \eqno{(6)}$$
This is the correct Fokker-Planck equation in the overdamped limit
and represents the Smoluchowski approximation to the original
eqn.(1). When the potential U(Q) is unbounded and positive, i.e,
$U(Q)\rightarrow \infty$ as $Q\rightarrow \pm\infty$, the system
evolves towards the stationary distribution $P_s(Q)$. This stationary
distribution is characterized by no net current flow and is given by
$$P_s (Q) = N e^{-\Psi (Q)},\eqno{(7)}$$
where N is normalization constant and
$$\Psi (Q)=  \int^Q\bigg \{ {U'(x)\Gamma (x)\over (T\Gamma_A+\overline T
\Gamma_B f(x))}+{(\overline T - T)\over\Gamma(x)} {\Gamma_A\Gamma_B
f'(x)\over (T\Gamma_A+\overline T \Gamma_B f(x))}\bigg \} dx .\eqno{(8)}$$
One can readily notice that in the equilibrium situation, i.e., when
$T = \overline T, P_s(Q)$ reduces to the correct equilibrium
distribution as given in eqn.(4).

To study the case of fluctuation-induced transport, we take a simple
case where both U(Q) and f(Q) are periodic functions and are
invariant under the same transformation $Q\rightarrow Q+\lambda$.
Now, the basic problem reduces to finding the mean velocity $<\dot
Q>$ of the subsystem given the shape of U(Q) and f(Q). Following the
procedure of refs.[9,12,13] closely, one can get the exact expression
for the averaged velocity
$$<\dot Q> = {[1-exp(-\delta)]\over \int^{\lambda}_0 dy \ exp[-\Psi
(y)]\int^{y+\lambda}_y{[\Gamma(x)]^2\over (T\Gamma_A+\overline T
\Gamma_B f(x))} \ exp [\Psi (x)] \ dx },\eqno{(9)}$$
where
$$\delta = \Psi (x) -\Psi (x+\lambda), \eqno{(10)}$$
and $\Psi$ is given by eqn.(8). It is easy to see from eqns.(9) and
(10), that in the equilibrium case, when the temperature difference
between the baths is zero $(T =\overline T)$, the current vanishes
indentically (since $\delta= 0$). Also, one can easily varify that
when the subsystem is coupled to a single bath, i.e., when either
$\Gamma_A$ or $\Gamma_B$ is zero, no net current is possible. It
should  be noted that the bath B which gives rise to space
dependent friction coefficient $\Gamma_B f(Q)$ plays a special role.
This can be noticed from the fact that if f(Q) is independent of Q
the induced current is zero. In the extreme case of high friction
limit $(\Gamma_A\rightarrow\infty \ or \ \Gamma_B\rightarrow\infty
)$ no net current is possible because the particle cannot execute the
Brownion motion.

There are two interesting special cases, one being the case of U(Q) =
0. It follows from eqn.(9) that even in the absence of external
potential fluctuation-induced currents are possible. In this case
direction of the current depends on the details of the function f(Q).
We see that the current will flow in one direction if $T < \overline T$
and in the opposite direction if $T > \overline T$. However, their
magnitudes are different. Thus, the system acts like a
Carnot engine,
 which  extracts work by making use of two thermal baths at
different temperatures. In the other case where amplitude modulations
in f(Q) and $f^{\prime}$(Q) are small compared to the amplitude modulation of
the potential U(Q), the second term in the eqn.(8) can be neglected.
In this particular limit the problem becomes equivalent to a particle
moving in a spatially varying temperature - field, namely, $T(Q) =
(T\Gamma_A+\overline T \Gamma_B f(Q))/\Gamma (Q)$. It is well-known
in earlier literature [12-14] that such a spatial modulation of
temperature-field can give induced currents. Moreover, the problem of
evolution of a particle in a spatially varying temperature field has a
fundamental consequence in relation to local versus global stability
criterion in statistical mechanics [15-16]. In
a thermal equilibrium state the Boltzmann factor gives a relative
occupation probability of states at different local stable points
without invoking the behaviour of the potential profile between the
states. However, in the presence of spatially varying
temperature-field, the relative stability (or occupation probability)
between different states depend sensitively on the intervening
potential and more importantly one can control the relative stability
among two different state by modfying the kinetics in the sparsely
occupied intervening state [14,16].

In conclusion, we have reinvestigated the recently proposed
self-consistent microscopic theory of fluctuation-induced transport.
We have pointed out errors in the earlier theory. We have obtained an
analytical expression for fluctuation-induced current in a
non-equilibrium situation and have discussed various cases of physical
interest.
Finally, we would like to mention that our expression for
the stationary distribution (or steady state) $P_s(Q)$ (eqn.(7)) is
not a local function of U(Q) and f(Q). In such a situation as
discussed above the relative stabiliy between two different local
states in U(Q) depends sensitively on the intervening behaviour of
U(Q) and f(Q). This can lead to interesting physics. For example, for
given external potential U(Q), as one varies physical parameters (in
the parameter space of T, $\overline T, \Gamma_A \ and \ \Gamma_B$)
we expect additional maxima and minima to appear in $P_s(Q)$. Thus we
can modify the stability properties of the subsystem. The qualitative
changes in the stationary state of the subsystem is reflected in
change in the behaviour of the extrema of the $P_s(Q)$. Moreover,
extrema in $P_s(Q)$ may have little or no relationship to the extrema
in the original potential U(Q). Each
new structure in $P_s(Q)$ may correspond to an entirely new
state of a subsystem. In the spirit of the well
known noise-induced phase transitions this
is equivalent to having hierarchy of phase
transitions, in a non-equilibrium system [17].
\vs
\noindent {\bf Acknowledgement}

The author thanks Mangal C. Mahato for several useful discussions
on this subject.
\vfill
\eject
\noindent {\bf References}

\item {[1]} R.P. Feynman, R.B. Leighton and M. Sands, The Feynman
Lectures in Physics (Addison - Wesley, Reading, 1966).
\item {[2]} M. Magnasco, Phys. Rev. Lett. {\bf 71}, 1477 (1993).
\item {[3]} J. Prost, J.F. Chauwin, L. Peliti and A. Ajdari, Phys.
Rev. Lett. {\bf 72}, 2652 (1994).
\item {[4]} C. Doering, W. Horsthemke and J. Riordan, Phys. Rev.
Lett. {\bf 72}, 2984 (1994).
\item {[5]}  R.D. Astumian and M. Bier, Phys. Rev. Lett. {\bf 72},
1766 (1994).
\item {[6]} M.M. Millonas and D.I. Dykman, Phys. Lett. A {\bf 183},
65 (1994).
\item {[7]} L.P. Faucheux, L.S. Bourdieu, P.D. Kaplan and A.J.
Libchaber, Phys. Rev. Lett. {\bf 74}, 1504 (1994).
\item {[8]} J. Rousselet, L. Salome, A. Ajdari and J. Prost, Nature
(London) {\bf 370}, 412 (1994).
\item {[9]} M.M. Millonas, Phys. Rev. Lett. {\bf 74}, 10 (1995).
\item {[10]} J.M. Jayannavar and M. C. Mahato, (unpublished).
\item {[11]} J.M. Sancho, M. San Miguel and D. Duerr, J. Stat. Phys.
{\bf 28}, 291 (1982).
\item {[12]} M. Buettiker, Z. Phys. B {\bf 68}, 161 (1987).
\item {[13]} N.G. van Kampen - IBM J. Res. Dev. {\bf 32}, 107 (1988).
\item {[14]} R. Landauer, Physica A {\bf 194}, 551 (1993).
\item {[15]} R. Landauer, Helv. Phys. Acta. {\bf 56,} 847 (1983).
\item {[16]} R. Landauer, Phys. Rev. A {\bf 12}, 636 (1975).
\item {[17]} W. Horsthemke and R. Lefever, Noise Induced Transitions
(Springer, New York, 1984).
\vfill
\eject
\end